\documentstyle[12pt]{article}

\newcommand{\reff}[1]{(\ref{#1})}
\newcommand{\bra}[1]{\langle{#1}|}
\newcommand{\ket}[1]{|{#1}\rangle}
\def\phan{\vphantom{\frac12}}

\def\non{\nonumber}
\def\al{\alpha}
\def\be{\beta}
\def\ga{\gamma}
\def\la{\lambda}
\def\La{\Lambda}
\def\D{{\cal D}}
\def\L{{\cal L}}
\def\de{\delta}
\def\cP{{\cal P}}
\author{ S. M. Klishevich%
\thanks{E-mail address: klishevich@th1.ihep.su} \\
        {\it  Institute for High Energy Physics
		} \\
        {\it  Protvino, Moscow Region, 142284, Russia
		}}
\title{Electromagnetic Interaction of Massive Spin-3 State from String
Theory.}
\begin{document}
\maketitle

\begin{abstract}
In the given work we study an interaction of second massive state of an open
boson string with the constant electromagnetic field. This state contains
massive fields with spins 3 and 1. Using the method of an open string BRST
quantization, we receive gauge-invariant lagrangian, describing the
electromagnetic interaction of these fields. From the explicit view of
transformations and lagrangian it follows that the presence of external
constant e/m field leads to the mixing of the given level states. Most
likely that the presence of the external field will lead to the mixing of
the states at other mass string levels as well.
\end{abstract}
\newpage

\section{Introduction.}

The problem of building the consistent interaction for high spin fields
has already sufficiently great history \cite{Dirac,Fierz,Rarita}. However,
and at present this problem is still far away from its completion.

The lagrangian description of free massive high spin fields was received by
Singh and Hagen \cite{Sing-Hagen} in the 70's.
Later it was realized \cite{Fang,Curt1} that for the covariant description
of massless fields, they must necessarily be gauge fields. For such fields
it was offered to enter an interaction as continuous deformation of the
gauge algebra and lagrangian \cite{Fronsdal-2}. Such entering of the
interaction is consistent since a continuous deformation of the gauge
algebra preserves a type of the equations defining mass shell. At the same
time it has been realized \cite{Zinoviev-2} that for massless fields of
spins $s\ge3/2$ in an asymptotically flat space-time it is impossible to
build a consistent "minimal" interaction with an Abelian vector field. The
same is valid and for the gravity interaction of fields with spins $s\ge2$.
It is possible to motivate the given statement as follows \cite{Zinoviev-2}:
A free gauge-invariant lagrangian in the flat space has a structure
$\L_0=\partial\Phi\partial\Phi$ with transformations $\de\Phi=\partial\xi$.
The Entering of the interaction means a replacement of the usual derivative
by a covariant one $\partial\to\D$. In this, the gauge invariance fails, and
a residual of type $\left[\D,\D\right]\D\Phi\xi={\cal R}\D\Phi\xi$ appears,
where ${\cal R}$ is a tensor of the tension of electromagnetic or gravity
field. In the case of the electromagnetic interaction for the fields with
spins $s\ge3/2$, one cannot cancel the residual with any changes of the
lagrangian and the transformations in the linear approximation. Therefore,
in this case such approximation is absent, but since any linear
approximation is independent of the presence in the system of any other
fields, this means that a whole theory of interaction does not exist either.
In the case of the gravity field a residual for the field spin 3/2 is
proportional to the gravity equations of motion:
$\de\L_0\sim i(\bar\psi^\mu\ga^\nu\eta)(R_{\mu\nu}-\frac12g_{\mu\nu}R)$. 
One can eliminate such residual by modifying the lagrangian and the
transformations. As a result, the theory of supergravity has appeared. For
fields with spins $s>2$ the residual contains terms proportional to Riemann
tensor $R_{\mu\nu\,\al\be}$. It is impossible to cancel such type terms  in
an asymptotically flat space. Hence, gravity interactions do not exist for
any massless fields with spin $s>3/2$.

It is possible to overcome similar difficulties in several ways. In the
case of the gravity interaction, one can consider fields in a constant
curvature space. Then the lagrangian of gravity would have an additional
term $\Delta\L=\sqrt{-{\rm det}\|g_{\mu\nu}\|}\la$, where $\la$ is a
cosmological constant. A modification of the lagrangian and the
transformations lead to a mixing of orders with the different derivative
numbers. This allows one to eliminate a residual with terms proportional to
$R_{\mu\nu\,\al\be}$. A complete theory will be represented as a series with
respect to inverse degrees of $\la$ \cite{Vas-DS-1,Vas-DS-2}. This means
non-analyticity of the theory with respect to $\la$ at zero, i.e.
the impossibility of a smooth transition to the flat space. Such a theory
was considered in papers \cite{Vas-DS-1,Vas-DS-2,Vasilev:IJMP95}.

 It is also possible to avoid these difficulties if one considers massive
high spin fields \cite{mass_spin}. But situation with the consistent
interaction of massive fields is significantly more complicated. The
classical description \cite{Sing-Hagen} is not gauge invariant. The Entering
of a "minimal" interactions lead to a change of the physical degree number
of freedom. This problem can be solved by entering non-minimal terms in the
interactions but, in a general case, it results in the loss of causality. 
If we have not a gauge invariance, then the universal principle that allows
one to get the consistent interaction of massive high spin fields is absent.

Early in the 80's the gauge invariant description of massive fields with
arbitrary spins \cite{Zinoviev-83} was offered. In this case one can
consider a consistent interaction as continuous deformation of the gauge
algebra similar to a massless case.

In a natural way, a gauge description of massive high spin fields 
appears in the frame of free string BRST quantization\footnote{%
One can also receive a gauge description of free massive fields of
arbitrary spins using dimensioned reductions
\cite{Sivakumar-PR85a,Sivakumar-PR85b}.} \cite{Siegel-NP86,Peskin-NP86}.
 At present this is practically the unique consistent theory describing
an interaction of high spin fields. However, in a general case, it is rather
difficult to get any particular information on separate spins from the
theory of interacting strings. But in some simple cases, such as a constant
electromagnetic field, it is possible. In \cite{argyres,argyresS1}
the authors generalized \cite{Siegel-NP86,Peskin-NP86} for the
case of  propagation of an open boson string in a homogeneous external e/m
field and obtained lagrangians which describe the consistent interaction of
massive fields of spins 2 and 1 with the constant electromagnetic fields.

Here we study an interaction of the second massive state of open boson
string with the constant electromagnetic field. This state contains massive
fields with spins 3 and 1. Using the BRST quantization method of the open
string, we get a gauge invariant lagrangian describing the electromagnetic
interaction of these fields. The action for gauge massive high spin fields
available from BRST formulation of free string \cite{Siegel-NP86} is
not canonical since a kinetic part contains cross terms. By means of
redefinitions of the fields, one can diagonalizate the action and consider
each state independently. In the presence of the external field, it
is already impossible to consider the states independently. From the
explicit form of the transformations and the lagrangian for the second mass
level, it follows that the presence of the constant e/m field leads to a
mixing of the states at the given level. Most likely that the presence of
the external field will lead to a mixing of the states at other mass levels
of the string as well.

\section{Electromagnetic interaction of high spin fields in context
boson string.}

Let us consider the action describing a propagation of the open boson string
with charges on the ends in a homogeneous electromagnetic field in
Minkow\-ski space ${\rm\bf M}_D$  in the conformal gauge
\cite{Fradkin-PL85,Nappi-NP87,argyres,argyresS1}
\begin{equation}
\label{String_in_cem}
S=\frac1{2\pi}\int\!\!d\tau\!\!\int\limits_0^\pi\!\!d\sigma\!
\left(\dot X_\mu\dot X^\mu - X'_\mu X'^\mu \right)
-\frac12\int\!\!d\tau d\sigma {\cal F}_{\mu\nu}\dot X^\mu X^\nu
(q_0\de(\sigma) + q_\pi\de(\sigma -\pi)),
\end{equation}
where the point and the prime denote a derivative with respect to the world
sheet coordinates of the string $\tau$ and $\sigma$, respectively. For the
tension coefficient the agreement $\alpha'=\frac12$ is accepted.

Equations of motion received from action \reff{String_in_cem} can be exactly
solved \cite{Nappi-NP87,argyres} which allows one to canonically quantize
the theory. Commutational relations for modes have following form:
\begin{equation}
\label{alg_in_cem}
\left[a_m^\mu,a_n^\nu\right]=H_m^{\mu\nu}\de_{m+n},
\end{equation}
with the usual hermitian conjugation ${a_n^\mu}^{\dag} =a_{-n}^\mu$. Above 
we used the following notations
\begin{eqnarray}
\label{ViewF}
H_m^{\mu\nu}&=&m\eta^{\mu\nu}+F^{\mu\nu},
\non\\
F^{\mu\nu}&=&\frac1{i\pi}\left({\rm arctanh}(\pi q_0{\cal F}) + 
\mbox{arctanh}(\pi q_\pi{\cal F})\right)^{\mu\nu}.
\end{eqnarray}

 Thus, the presence of the constant electromagnetic field leads to the
deformation of the infinite Heisenberg algebra.

The Fok space, on which the representation of algebra \reff{alg_in_cem} is
realized, is built in the same manner as in the absence of the external
field. The vacuum vector is defined by the relations
$$
a_n^\mu|0\rangle=0,\quad\forall\;n<0,
$$
and any vector of the Fok space has the form
$$
|\Phi\rangle=\sum_k\sum_{\{n_i\}}\Phi_{\mu_1...\mu_k}^{(n_1...n_k)}(x)
a^{\mu_1}_{-n_1}...a^{\mu_k}_{-n_k}|0\rangle, 
\quad{\rm где }\quad n_i>0.
$$
The coefficient functions $\Phi_{\mu_1...\mu_k}^{(n_1...n_k)}(x)$ are
tensor fields on the space ${\rm\bf M}_D$.

The theory with action \reff{String_in_cem}, as in the case with the free
string, contains constraints expressed in terms of Virasoro operators at the
quantum level. In spite of the presence of the external field, the Virasoro
operators corresponding to action \reff{String_in_cem} have the usual form
\begin{equation}
\label{Virasoro}
L_m=\frac1{4\pi}\int\limits_{-\pi}^{\pi}e^{-im\sigma}
\left(\dot X + X'\right)^2d\sigma=
\frac12\sum_p:\!a_{m-p}\cdot a_p\!:.
\end{equation}
They form the same algebra but with other central charge \cite{Nappi-NP87}
$$
\left[L_m,L_n\right]=(m-n)L_{m+n}
+ \left\{\frac{D}{12}(m^3-m) - \frac12mF_{\mu\nu}^2
\right\}\de_{m+n}
$$
These relations can be reduced to the same form as for the free string.
For that it is necessary to redefine the operator $L_0$ including into its
definition the term ${}-\frac14F_{\mu\nu}^2$. This lead only to the
redefining of a normal ordering constant\footnote{Which we will include 
into the defenition of $L_0$ as well.}.

A subspace of physical states is defined by means of the following
conditions:
\begin{equation}
\label{L-cond}
L_n|\Psi\rangle=0, \quad\forall n\ge0.
\end{equation}

For the free string, a representation of the zero mode is, as usual, chosen
in the form of the momentum operator $a_0^\mu=p^\mu=i\partial^\mu$ acting on
the space of functions $C^{\infty}({\rm\bf M}_D)$. In such representation,
one can get from \reff{L-cond} the differential equations for coefficient
function of the physical states. As is well known that the conditions
\reff{L-cond} contain the equations for the massive states with an arbitrary
integer spin.

Let us consider the commutation relations for the zero modes in the presence
of the external electromagnetic field
$$
[a_0^\mu,a_0^\nu]=F^{\mu\nu}.
$$
In the constant field the particular view of $F^{\mu\nu}$ \reff{ViewF} does
not matter\footnote{It is possible to take $F^{\mu\nu}$ as an arbitrary
uneven function of a constant antisymmetric matrix.}, therefore, we "will
forget" about the origin of algebra \reff{alg_in_cem} and interpret
the antisymmetric tensor $F^{\mu\nu}$ as a tension of the electromagnetic
field. We consider that the coupling constant and the imaginary unity are
included in the definition of the tension tensor. Then $a_0^\mu$ can be
interpreted as a covariant momentum operator
$$
a_0^\mu={\cal P}_\mu=i\D^\mu,
$$
where $\D^\mu$ is $U(1)_{em}$ covariant derivative. Under such
interpretation, conditions \reff{L-cond} define the equations of motion
for the physical states with arbitrary spins in the constant electromagnetic
field. But it is necessary to notice that the obtaining of a lagrangian
formulation, which can give all these equations in such form, is
sufficiently difficult.

In \cite{Siegel-NP86,Peskin-NP86} gauge invariant equations for the free
string states coming from the requirement of BRST invariance was proposed.
In \cite{argyres} the author showed that in the case of homogeneous
electromagnetic field, the BRST charge is built in the same
manner as in the case of the free string.

Let us increase the above considered Fok space by entering an infinite set
of "ghost" and "antighost" operators $c_n$ and $b_n$ with anticommutation
relations
\begin{eqnarray*}
&[c_m,b_n]_{+}=\de_{m+n},&\\
&[c_m,c_n]_{+}=[b_m,b_n]_{+}=0&
\end{eqnarray*}
and with the usual hermitian conjugation $c_n^{\dag}=c_{-n}$,
$b_n^{\dag}=b_{-n}$. A "ghost" vacuum is defined as follows
\begin{eqnarray*}
&c_n|0\rangle_{gh}=0,&\quad\forall n>1,
\\
&b_n|0\rangle_{gh}=0,&\quad\forall n>-2.
\end{eqnarray*}
A complete vacuum is defined as a tensor product of the two vacuum vectors
$$
|0\rangle=|0\rangle_{X}\otimes|0\rangle_{gh},
$$
where $|0\rangle_X$ is the boson vacuum vector defined by conditions
\reff{alg_in_cem}. The complete vacuum is $sl(2,R)$ invariant.

In the theory of the free string BRST quantizing the states are classified
by means of the "level" operator
$$
{\cal N}=\sum_{p=1}^\infty\left(a_{-p}\cdot a_p
+ pc_{-p}b_p\right) 
$$
and the operator of a "ghost" number \cite{Kato:NP83,Kugo:79}
$$
{\cal N}_{gh}=\frac12[c_0,b_0] + \sum_{p=1}^\infty\left(
c_{-p}b_p - b_{-p}c_p\right).
$$
Physical states are eigenvectors of these operators.The vacuum is the
physical state with "ghost" number $-3/2$. In the theory with an external
field we shall use the same states as in the free theory, but the states
will not be eigenvectors of the operators ${\cal N}$ and ${\cal N}_{gh}$
already because of the presence in the commutation relations
\reff{alg_in_cem} of the tension tensor of the electromagnetic field.

For the feature of the physical states in the presence of the external field
we shall use the same terms as for the free string. We shall assume that its
is defined in the limit $F_{\mu\nu}\to0$.

Both for the free string and for the one in a constant Abelian field, the
BRST operator is defined as
\begin{equation}
\label{BRST-charge}
Q_{BRST}=\sum_{-\infty}^{\infty}
:\left(L_p^X + \frac12L_p^{gh}\right)c_{-p}:,
\end{equation}
where operators $L_m^X$ are defined by expression \reff{Virasoro},
$L_m^{gh}=\sum_p(m-p)b_{p+m}c_{-p}$ and the normal ordering constant is
included into the definition of $L_0$.  As for the free string, this
Hermitian operator is nilpotent in the space with dimensionality 26. 
Equations \reff{L-cond} are equivalent to the requirement of the BRST
invariance for physical states
\begin{equation}
\label{Q-eq}
Q_{BRST}\ket{\Psi}=0.
\end{equation}
The physical states $\ket{\Psi}$ and $\ket{\Psi}+Q_{BRST}\ket{\La}$ are
equivalent as a consequence of the nilpotency of the BRST operator. Here 
the "ghost" number of the vector $\ket{\La}$ is on the unity less than
the number of $\ket{\Psi}$. This means that equations \reff{Q-eq} are
gauge invariant.

Equations \reff{Q-eq} can be derived from the lagrangian
\begin{equation}
\label{<Q>}
\L=\bra{\Psi}Q_{BRST}\ket{\Psi},
\end{equation}
where $\bra{\Psi}=\ket{\Psi}^{\dag}$, $\bra{0}=\ket{0}^{\dag}$ and an
uncertainty in the determination of the vacuum average is fixed by taking
the agreement
$$
\langle0|c_{-1}c_0c_{1}|0\rangle=1.
$$

Obviously, that this lagrangian is invariant under the gauge transformations
\begin{equation}
\de\ket{\Psi}=Q_{BRST}\ket{\La}.
\end{equation}

Thus, the BRST formulation of the open boson string allows one to obtain 
the gauge invariant lagrangians describing the massive fields with 
arbitrary integer spins.

\section{The second massive string state.}

In \cite{argyres} the authors studied the first massive state of the
open boson string and using the above method, they proposed the following
lagrangian describing a propagation of the massive spin 2 field in a
constant electromagnetic field 
\begin{eqnarray}
\label{L-sp2}
\L&\!=\!&{}-\frac12\bar\phi(\cP H_1\cP+2)\phi
-\frac12\bar B H_2\cdot(\cP H_1\cP+2)B
-\bar BH_2 F B
\non\\&&{}
+\mbox{tr}\left[\bar H_1\bar h H_1\cdot(\cP H_1\cP+2)h\right]
+4\mbox{tr}\left[\bar H_1 \bar h H_1 F h\right]
\\\non&&{}
+\frac12\left(2\cP\bar H_1\bar h+\bar H_2\bar B-\cP\bar \phi\right)
\cdot H_1\cdot\left(2\cP H_1 h+H_2 B-\cP\phi\right)
\\\non&&{}
-\frac14\left(
 \mbox{tr}\left[\bar H_1\bar hH_1\right]
 +\cP\bar H_2 \bar B -3\bar \phi \right)
\left(
 \mbox{tr}\left[H_1h \bar H_1\right]
 +\cP H_2 B -3\phi \right),
\end{eqnarray}
which is invariant with respect to the gauge transformations
\begin{eqnarray}
\label{var-sp2}
 \de h_{\mu\nu}&=&\cP_\mu\xi^1_\nu+\cP_\nu\xi^1_\mu-\frac12g_{\mu\nu}\xi^0,
\non\\
 \de B_\mu&=&-\cP_\mu\xi^0+2\left(H_1\cdot\xi^1\right)_\mu,
\\\non
 \de \phi&=& 2\left(\cP\cdot H_1\cdot\xi^1\right)-3\xi^0
\end{eqnarray}
Here, the bar means the complex conjugation, $h_{\mu\nu}$ is a symmetrical
tensor. We used the matrix notations as well
$\left(\cP H_1\cP\right)=\cP_\mu H_1^{\mu\nu}\cP_\nu$, 
$\left(\cP H_1 h\right)_\mu=\cP_\al H_1^{\al\be}h_{\be\mu}$ and etc.

In \cite{argyres} it has been also proved that the propagation of the
massive spin 2 field in the constant field described by  lagrangian
\reff{L-sp2} is causal. This allows us to hope that \reff{<Q>} gives
consistent description of the interaction with the homogeneous
electromagnetic field for the massive ones of arbitrary integer
spins.

The second massive state of the open boson string contains the massive
fields with spins 1 and 3. This means that the average \reff{<Q>} of the
BRST operator with respect to the present state gives us a gauge invariant
lagrangian describing the interaction of these fields with the constant
electromagnetic field.

The second massive string level state is defined by the conditions
$$
{\cal N}\ket{\Psi,3}\;\stackrel{{F\to0}}{\longrightarrow}\;
2\ket{\Psi,3},\quad
 {\cal N}_{gh}\ket{\Psi,3}=\ket{\Psi,3}
$$
and has the following form
\begin{eqnarray}
|\Phi,3\rangle&=&{}
\{
\Phi_{\mu\nu\al} a_{-1}^\mu a_{-1}^\nu a_{-1}^{\al} c_1
 + \left(h_{\mu\nu}+A_{\mu\nu}\right) a_{-2}^\mu a_{-1}^\nu c_1
 + B_\mu a_{-3}^\mu c_1
\non\\\non&&{}
 + A_\mu a_{-1}^\mu c_{-1}
 + \varphi c_{-2} + \sigma b_{-2}c_1c_{-1}
 + \omega_{\mu\nu} a_{-1}^\mu a_{-1}^\nu c_0
 + \theta_\mu a_{-2}^\mu c_0
\\&&{}
 + \xi_\mu a_{-1}^\mu b_{-2} c_0 c_1
 + \ga b_{-3} c_0 c_1
\}
|0\rangle.
\end{eqnarray}
Here $A_{\mu\nu}$ is an antisymmetric and
$\Phi_{\mu\nu\al},h_{\mu\nu},\omega_{\mu\nu}$ are symmetrical tensor
fields.

The gauge transformation for this states in terms of the BRST operator is
$$
\de|\Phi,3\rangle=Q_{BRST}|\La,3\rangle,
$$
where 
\begin{eqnarray*}
|\Lambda,3\rangle&=&
\{\Lambda_{\mu\nu} a_{-1}^\mu a_{-1}^\nu
 + x_\mu a_{-2}^\mu
 + \lambda_\mu a_{-1}^\mu b_{-2} c_1
 + \eta b_{-3} c_1 + v b_{-2} c_0
\}|0\rangle.
\end{eqnarray*}
Here $\Lambda_{\mu\nu}$ is an arbitrary symmetrical tensor. Hereafter we
shall not distinguish the upper and lower indexes.

In terms of the coefficient functions, the gauge transformations take the
following form
\begin{eqnarray}
\label{trans0}
\delta \Phi _{\alpha \beta \gamma }&=&i
{\cal D}_{(\gamma }\Lambda _{\alpha \beta )}+\frac 12g_{(\alpha \beta
}\lambda _{\gamma )}, 
\non\\\phan{}
 \delta h_{\alpha \beta }&=&i
{\cal D}_{(\alpha }x_{\beta )}+i{\cal D}_{(\alpha }\lambda _{\beta
)}+2H_{1(\alpha |\gamma }\Lambda _{\gamma |\beta )}+g_{\alpha \beta }\eta,
\non\\\phan{}
 \delta A_{\alpha \beta }&=&i
{\cal D}_{[\alpha }\lambda _{\beta ]}-i{\cal D}_{[\alpha }x_{\beta
]}+2F_{[\alpha |\gamma }\Lambda _{\gamma |\beta ]}, 
\non\\\phan{}
 \delta b_\alpha &=&i
{\cal D}_\alpha \eta +H_{2\alpha \beta }x_\beta +H_{1\alpha \beta }\lambda
_\beta , 
\\\phan{}
 \delta A_\alpha &=&i2H_{1\beta \gamma }
{\cal D}_\beta \Lambda _{\gamma \alpha }+H_{2\alpha \beta }x_\beta
+3\lambda_\alpha, 
\non\\\phan{}
 \delta \varphi &=&iH_{2\alpha \beta }
{\cal D}_\alpha x_\beta +\bar H_{1\alpha \beta }H_{1\beta \gamma
}\Lambda _{\alpha \gamma }+4v+5\eta , 
\non\\\phan{}
 \delta \sigma&=&-iH_{1\alpha \beta}{\cal D}_\alpha \lambda _\beta
 +2v-4\eta , 
\non\\\non&&\\
\delta \omega _{\alpha \beta }&=&-\frac 12H_{1\gamma \delta }
{\cal D}_{\gamma \delta }^2\Lambda _{\alpha \beta }+2H_{1(\alpha |\gamma
}\Lambda _{\gamma |\beta )}+\frac 12g_{\alpha \beta }v, 
\non\\\phan{} \delta
\theta
_\alpha &=&-\frac 12H_{1\beta \gamma }
{\cal D}_{\beta \gamma }^2x_\alpha +i{\cal D}_\alpha v+H_{2\alpha \beta
}x_\beta , 
\\\non\phan{}
 \delta \chi _\alpha \,&=&\frac 12H_{1\beta \gamma }
{\cal D}_{\beta \gamma }^2\lambda _\alpha +i{\cal D}_\alpha v-H_{2\alpha
\beta }\lambda _\beta , 
\\\non\phan{}
 \de\ga&=&\frac12H_{1\al\be}\D_{\alpha\beta}^2\eta -2\eta + v.
\end{eqnarray}
The fields $\omega_{\alpha\beta},\theta_\alpha,\chi_\alpha,\gamma$ are
auxiliary since they are coefficient functions at the combinations of
operators containing $c_0$ but the BRST operator has the structure
$$
Q_{BRST}=-\frac12\D^2c_0 + ...,
$$
where the dots are marked terms which do not exceed the first order for the
derivatives. Therefore, the auxiliary fields have an algebraic equations
on mass surfaces. Having solved this equation, one can to exclude all the
auxiliary fields from our consideration. However, it is the auxiliary
fields that allow one to write down a lagrangian for the second massive
state in the most compact form.

From the transformations \reff{trans0} it is not difficult to see that the
scalar parameter $v$ appears only as a translation. This means that we can
discard one scalar field, for instance $\sigma$, using this parameter.
Herewith the gauge transformations are changed only for
field
$\varphi$
$$
\de_{new}\varphi=
iH_{2\al\be}\D_\al x_\be + 2iH_{1\al\be}\D_\al\la_\be
 + \bar H_{1\al\be}H_{1\be\ga}\La_{\al\ga} + 13\eta
$$
 and for the auxiliary fields.

After excluding the field $\sigma$, the gauge lagrangian of the second
massive state will take the form
\begin{eqnarray}
\label{Lwith}
\L&=&\langle\Phi,3|Q_{BRST}|\Phi,3\rangle\biggr|_{\sigma\to0}
\non\\\phan&=&
3\bar H_{1\alpha \mu }\bar H_{1\beta \nu }\bar H_{1\gamma
\rho }\bar H_{1\delta \lambda }{\cal D}_\delta \Phi _{\alpha \beta
\gamma }{\cal D}_\lambda \bar \Phi _{\mu \nu \rho } 
\\\non\phan&&{}
 +18\bar\Phi_{\alpha
\beta \gamma }H_{1\alpha \delta }H_{1\beta \rho }H_{1\rho \mu }H_{1\gamma
\nu }\Phi _{\delta \mu \nu }-6\bar \Phi _{\alpha \beta \gamma }H_{1\alpha
\delta }H_{1\beta \mu }H_{1\gamma \nu }\Phi _{\delta \mu \nu } 
\\\non\phan&&{} 
-6\bar H_{1\mu \beta }\bar \Phi _{\alpha \beta \gamma }H_{1\alpha
\delta
}H_{1\gamma \nu }{\cal D}_\nu \omega _{\delta \mu }i+3\bar \Phi _{\alpha
\beta \gamma }H_{1\alpha \delta }\xi _\delta H_{1\beta \mu }\bar H%
_{1\mu \gamma } 
\\\non\phan&&{}
 +\frac 12
\bar H_{1\gamma \nu }\bar H_{1\beta \mu }\bar H_{2\alpha
\delta }{\cal D}_\gamma h_{\alpha \beta }{\cal D}_\nu \bar h_{\delta \mu
}+\bar h_{\alpha \beta }H_{2\beta \gamma }H_{2\gamma \delta }\bar H%
_{1\mu \alpha }h_{\delta \mu } 
\\\non\phan&&{}
 -\bar h_{\alpha \beta}H_{2\beta\gamma}h_{\gamma \delta }F_{\delta \mu }
\bar H_{1\mu \alpha }+\frac 12\bar H_{2\alpha \delta
}\bar H_{1\beta \mu }\bar H_{1\gamma \nu }{\cal D}_\gamma 
A_{\alpha \beta}
{\cal D}_\nu \bar h_{\delta \mu } 
\\\non\phan&&{}
 +\bar h_{\alpha \beta}H_{2\beta\gamma}H_{2\gamma \delta }A_{\delta \mu }
\bar H_{1\mu \alpha }-\bar h_{\alpha \beta }H_{2\beta \gamma
}A_{\gamma \delta }F_{\delta \mu }\bar H_{1\mu \alpha } 
\\\non\phan&&{}
-2\bar h_{\alpha \beta }H_{2\beta \gamma }H_{1\gamma \delta }
\omega_{\delta \mu}
\bar H_{1\mu \alpha }-\bar H_{2\alpha \beta }\bar h_{\beta
\gamma }H_{1\gamma \delta }{\cal D}_\delta \theta _\alpha i 
\\\non\phan&&{}
 +\bar H_{1\alpha \beta }\bar h_{\beta \gamma }H_{2\gamma \delta
}{\cal D%
}_\delta \xi _\alpha i+\bar h_{\alpha \beta }H_{1\beta \gamma
}\bar H%
_{2\gamma \alpha }\gamma  
\\\non\phan&&{}
 -\{\frac 12H_{1\alpha \delta}H_{1\gamma\nu}H_{2\beta \mu }
{\cal D}_\gamma \bar A_{\alpha \beta }{\cal D}_\nu h_{\delta \mu }+\bar
A_{\alpha \beta }H_{2\beta \gamma }H_{2\gamma \delta }h_{\delta \mu }
\bar H_{1\mu \alpha } 
\\\non\phan&&{}
 -\bar A_{\alpha \beta }H_{2\beta\gamma
}h_{\gamma \delta }F_{\delta \mu }
\bar H_{1\mu \alpha }+\frac 12\bar H_{1\gamma \nu
}\bar H%
_{2\alpha \mu }\bar H_{1\beta \delta }{\cal D}_\gamma A_{\alpha
\beta }%
{\cal D}_\nu \bar A_{\delta \mu } 
\\\non\phan&&{}
 +\bar A_{\alpha \beta}H_{2\beta\gamma}H_{2\gamma \delta }A_{\delta \mu }
\bar H_{1\mu \alpha }-\bar A_{\alpha \beta }H_{2\beta \gamma
}A_{\gamma \delta }F_{\delta \mu }\bar H_{1\mu \alpha } 
\\\non\phan&&{}
-2\bar A_{\alpha \beta }H_{2\beta \gamma }H_{1\gamma \delta }
\omega_{\delta \mu}
\bar H_{1\mu \alpha }+\bar H_{2\alpha \beta }\bar A_{\beta
\gamma }H_{1\gamma \delta }{\cal D}_\delta \theta _\alpha i 
\\\non\phan&&{}
 + \bar H_{1\alpha \beta }\bar A_{\beta \gamma }H_{2\gamma \delta
}{\cal D}_\delta \xi _\alpha i\}
+\bar A_{\alpha \beta }H_{1\beta \gamma}\bar H_{2\gamma \alpha }\gamma  
\\\non\phan&&{}
 +\frac 12
\bar H_{3\alpha \gamma }\bar H_{1\beta \delta }{\cal D}_\beta
b_\alpha {\cal D}_\delta \bar b_\gamma +\bar b_\alpha H_{3\alpha \beta
}H_{2\beta \gamma }b_\gamma
\\\non\phan&&{}
 -\bar b_\alpha H_{3\alpha \beta }H_{2\beta\gamma }\theta _\gamma  
 +\bar b_\alpha H_{3\alpha \beta}H_{1\beta \gamma}\xi _\gamma
 +\bar b_\alpha H_{3\alpha \beta }{\cal D}_\beta \gamma i
\\\non\phan&&{}
 -\bar \varphi\{H_{1\alpha \beta }
{\cal D}_\alpha \xi _\beta i+4\gamma \}-\{6\bar H_{1\alpha \beta}\bar
\omega _{\beta \gamma }H_{1\gamma \delta }H_{1\mu \nu }{\cal D}_\mu \Phi
_{\alpha \delta \nu }i 
\\\non\phan&&{}
 +2\bar \omega _{\alpha \beta }H_{1\beta\gamma
}H_{2\gamma \delta }\left(h_{\delta \mu }+A_{\delta \mu }\right)
\bar H_{1\mu \alpha }-2\bar H_{1\alpha \beta }\bar \omega
_{\beta \gamma }H_{1\gamma \delta }{\cal D}_\delta A_\alpha i 
\\\non\phan&&{}
-4\bar \omega _{\alpha \beta }H_{1\beta \gamma }\omega _{\gamma \delta }
\bar H_{1\delta \alpha }\}-H_{1\alpha \beta }\left({\cal
D}_\alpha A_{\gamma \beta }
 +{\cal D}_\alpha h_{\gamma \beta }\right)\bar H_{2\gamma \delta }\bar
\theta_\delta i
\\\non\phan&&{}
+\bar \theta _\alpha H_{2\alpha \beta }H_{1\beta \gamma }A_\gamma  
-\bar\theta_\alpha H_{2\alpha \beta }H_{3\beta \gamma }b_\gamma
 +2\bar \theta _\alpha H_{2\alpha \beta }\theta _\beta
\\\non\phan&&{} 
 -\frac 12\bar H_{1\beta \delta }\bar H_{1\alpha \gamma }{\cal D}_\beta
A_\alpha {\cal D}_\delta \bar A_\gamma  
 - \bar A_\alpha H_{1\alpha\beta}H_{2\beta \gamma }A_\gamma
\\\non\phan&&{}
 + 2H_{1\alpha \beta}{\cal D}_\alpha \omega _{\beta \gamma }\bar H_{1\gamma
\delta }\bar A_\delta i
 + \bar A_\alpha H_{1\alpha \beta }H_{2\beta \gamma }\theta_\gamma 
 - 3\bar A_\alpha H_{1\alpha \beta }\xi _\beta
\\\non\phan&&{}
 + H_{2\alpha \beta}\left({\cal D}_\alpha A_{\beta \gamma }+{\cal D}_\alpha
h_{\beta \gamma }\right)\bar H_{1\gamma \delta }\bar \xi _\delta i 
 +3\bar \xi_\alpha H_{1\alpha \beta }\Phi _{\beta \gamma \mu }H_{1\gamma
  \delta }\bar H_{1\delta \mu }
\\\non\phan&&{}
+\bar \xi _\alpha H_{1\alpha \beta }H_{3\beta
\gamma }b_\gamma -\bar \xi _\alpha H_{1\alpha \beta }{\cal D}_\beta
\varphi i 
 +4\bar \xi _\alpha H_{1\alpha \beta }\xi _\beta -3\bar \xi_\alpha
H_{1\alpha \beta }A_\beta
\\\non\phan&&{}
 +\bar \gamma \{H_{3\alpha \beta }
{\cal D}_\alpha b_\beta i 
 +H_{2\alpha \beta }\left(h_{\beta\gamma}+A_{\beta\gamma}\right)\bar
H_{1\gamma \alpha }-4\varphi +6\gamma\}.
\end{eqnarray}

Varying this lagrangian with respect to auxiliary fields, it is not
difficult to make sure that the equations of motion for them are really
algebraic. Using this equations, one can exclude the auxiliary fields from
the lagrangian \reff{Lwith}. We have placed the obtained lagrangian
\reff{Lful} into Appendix because it is rather long. It is necessary to note
that the received result is complete with respect to the tension of the
electromagnetic field.

 The lagrangian \reff{Lful} for the interacting fields of spins 3 and 1
is non-canonical, since the free part of the lagrangian contains the cross
kinetic terms. To diagonalize them, it is necessary to do the following
replacement of the variables
\begin{eqnarray}
\label{Subst}
\Phi_{\al\be\ga}&\to&\Phi_{\al\be\ga}
 + \frac1{416}g_{(\al\be}\left(13A_{\ga)} + 3 b_{\ga)}\right),
\non\\\non
h_{\al\be}&\to& h_{\al\be} - \frac34g_{\al\be}\varphi,
\non\\
A_\al &\to& \frac1{208}\left(13 A_\al + 3 b_\al\right)
 + 3 \Phi_{\al\be\be},
\\
b_\al &\to& \frac1{16}\left(b_\al - A_\al\right),
\non\\\non
\varphi&\to& 2 h_{\al\al} - 3 \varphi.
\end{eqnarray}
After such changing, the fields $\Phi_{\al\be\ga},h_{\al\be},b_\al,\varphi$
will describe the pure massive spin 3 (see, for instance,
\cite{mass_spin,Siegel-NP86}). The gauge transformations for these fields
will
acquire the form
\begin{eqnarray}
\label{Dtrans3}
\de\Phi_{\al\be\ga}&=&i\D_{(\al}\tilde\La_{\be\ga)}
+ \frac{i}{13}g_{(\al\be|}F_{\mu\nu}\D_\mu\tilde\La_{\nu|\ga)}
- \frac{i}{5408}g_{(\al\be}F_{\ga)\mu}\D_\mu
\left(3\tilde u + 13\tilde\eta\right)
\non\\&&
+ \frac1{52}g_{(\al\be}\left(H_{4\ga)\mu}\tilde x_\mu
 - F_{\ga)\mu}\tilde\la_\mu\right)
\non\\
\de h_{\al\be}&=& i\D_{(\al}\tilde x_{\be)}
 + 2H_{1(\al|\mu}\tilde\La_{\mu|\be)}
+ \frac{43}{52}g_{\al\be}\tilde u
 - \frac1{8}g_{\al\be}\Bigl(3iF_{\mu\nu}\D_\mu\tilde x_\nu
\non\\&&
+ iF_{\mu\nu}\D_\mu\tilde\la_\nu
- 2F_{\mu\nu}F_{\nu\ga}\tilde\La_{\mu\ga}
+ \frac1{208} F_{\mu\nu}^2\left(3\tilde u + 13 \tilde\eta\right)\Bigr)
\\
\de b_\al&=&i\D_\al\tilde u + 24\tilde x_\al
 - 2iF_{\mu\nu}\D_\mu\tilde\La_{\nu\al}
+ \frac{i}{208}F_{\al\mu}\D_\mu\left(3\tilde u + 13 \tilde\eta\right)
\non\\&&
+ \frac12F_{\al\mu}\left( 25\tilde x_\mu + \tilde\la_\mu\right)
\non\\
\de\varphi &=& \tilde u
-\frac{i}6F_{\al\be}\left(3\D_\al\tilde x_\be+\D_\al\tilde\la_\be\right)
+\frac13F_{\al\be}F_{\be\ga}\tilde\La_{\al\ga}
\non\\\non&&
- \frac1{1248}F_{\al\be}^2\left(3\tilde u + 13\tilde\eta\right).
\end{eqnarray}
The gauge parameters are redefined as follows:
\begin{eqnarray*}
\tilde\La_{\al\be}&=&\La_{\al\be} -\frac1{26}g_{\al\be}\La_{\ga\ga},
\\
\tilde x_\al&=&x_\al + \la_\al,
\\
\tilde u &=& \La_{\al\al}+13\eta,
\\
\tilde\la_\al &=&\la_\al - x_\al,
\\
\tilde\eta&=&\La_{\al\al} - 3 \eta.
\end{eqnarray*}
In turn, the fields $A_{\al\be},A_\al$ will relate to the massive spin 1
with
the gauge transformations
\begin{eqnarray}
\label{Dtrans1}
\de A_{\al\be}&=& i\D_{[\al}\tilde\la_{\be]}
 + 2 F_{[\al|\ga} \tilde\La_{\ga|\be]},
\non\\
\de A_\al&=& i\D_\al\tilde\eta + 8 \tilde\la_\al
 - 2iF_{\be\ga}\D_\be\tilde\La_{\ga\al}
+ \frac{i}{208}F_{\al\be}\D_\be\left(3 \tilde x + 13 \tilde\eta\right)
\non\\&&
 - \frac12F_{\al\be}\left(7\tilde x_\be - \tilde\la_\be\right).
\end{eqnarray}
Under such substitution lagrangian \reff{Lful} increases approximately
three times. For this reason we do not display it here.

From transformations \reff{Dtrans3} and \reff{Dtrans1} it is clear that in
the presence of the constant electromagnetic field a mixing of the states
with spins 3 and 1 occurs, i.e. in the presence of the interaction we cannot
any longer consider these states independently from each other. It would
appear reasonable that the presence of the constant Abelian field leads to
the mixing of the states for any other higher mass level of the open boson
string as well.

At the end of this section we notice that the massive spin 1 state appears
at the second massive string level in the non-standard form (see also
\cite{Siegel-NP86}). The free lagrangian corresponding to a noncharged
massive spin 1 can be reduced to the form
\begin{equation}
\label{Lspin1}
\L_{S=1} =\frac1{12}{\cal H}_{\al\be\ga}{\cal H}^{\al\be\ga}
 -\frac14{\cal H}_{\al\be}{\cal H}^{\al\be}
+ \frac{m}{2}{\cal H}_{\al\be}A^{\al\be}
 - \frac{m^2}{4}A_{\al\be}A^{\al\be}
\end{equation}
by a change of field normalizations. Here $ {\cal H}_{\al\be\ga}=
\partial_\al A_{\be\ga} + \partial_\be A_{\ga\al} + \partial_\ga
A_{\al\be}$, ${\cal H}_{\al\be}=\partial_\al A_\be - \partial_\be A_\al$ 
and the dependency on the dimensional parameter was restored as well. This
lagrangian is invariant under the gauge transformations
\begin{eqnarray*}
\de A_{\al\be}&=&\partial_{[\al}\la_{\be]}
\\
\de A_\al &=&\partial_\al\eta  + \la_\al.
\end{eqnarray*}
The field $A_\al$ is the Goldstone one for the antisymmetric tensor.
Herewith the gauge algebra is reducible as in the case with the
vector-tensor models with the topological coupling
\cite{Allen:MPL91,Lahiri:MPL93}. However, unlike the last mentioned, it is
obvious that the form of the lagrangian \reff{Lspin1} does not depend on the
dimensionality of the space-time.

\section{About high spin interactions in non-critical dimensionality.}

Lagrangians \reff{L-sp2} and \reff{Lful} obtained in the context of
boson string are gauge invariant in the space-time of critical
dimensionality 26 only. However, this does not mean that high spin
interaction is consistent for the critical dimensionality only. So, for
instance, in \cite{mass_spin} using the Noether procedure a complete gauge
invariant lagrangian describing interaction of massive spin 2
field with the constant electromagnetic field in the space-time of any
dimensionality was derived. Similarly, one can consider the massive spin 3
field as well. Besides, it is possible to get an interaction of the high
spin fields with the constant field in a non-critical dimensionality from
the BRST quantization of "massive" strings
\cite{Fujikawa:PL89,Hasiewicz:NP96}.

The transition from the usual boson string to the "massive" one is realized
by the following modification of action \reff{String_in_cem} (refer to
\cite{Hasiewicz:NP96})
$$
\tilde S= -\frac{\al'}{2\pi}\int\!\!\sqrt{-g}d^2\xi g^{ab}
\partial_aX^\mu\partial_bX^\nu\eta_{\mu\nu}
 - \frac{\be}{2\pi}\int\!\!\sqrt{-g}d^2\xi
\left(g^{ab}\partial_a\varphi\partial_b\varphi + 2R_g\varphi\right).
$$
Here $\be$ is an arbitrary non-negative parameter.

Under quantization, such modification leads to the expansion of the Fok
space of the usual string by adding a infinite set of the scalar operators
$b_m$ with commutation relations
\begin{equation}\non
\left[b_m,b_n\right]=m\de_{m+n}
\end{equation}
and to the change of the definition for the Virasoro operators
\begin{equation}
\label{Vir+b}
L^X_m=\frac12\sum:\!a_{m+p}\cdot a_{-p}\!:
+ \frac12\sum:\!b_{m+p}b_{-p}\!: + 2\sqrt{\be}imb_m + 2\be\de_m.
\end{equation}
We are always able to include the last term to the definition of $L_0$.
Besides, a new constraint on the physical states appears
$$
\left(b_0 - q\right)\ket{\Psi}=0,
$$
where $q$ is an arbitrary real parameter.
This requirement leads to a shift of the theory spectrum. To see it we write
down the condition $L_0\ket{\Psi}$ in the absence of the external field
$$
\left(-\frac12 M^2
 + \frac12\sum\limits_{p=1}^\infty\left(a_{-p}\cdot a_p +
b_{-p}b_p\right) + \frac12 b_0{}^2 - 1
\right)\ket{\Psi}=0,
$$
i.e. the mass of any string level gets difference $q^2$.
Hence it is clear that if $q^2\geq2$ the tahyons will be absent in the
spectrum.

Operators \reff{Vir+b} satisfy the Virasoro algebra with the other
central charge
\begin{equation}
\label{alg_L_b}
\left[L_m,L_n\right]=(m-n)L_{m+n}+\frac{D+1+48\be}{12}(m^3-m)\de_{m+n}.
\end{equation}
From these relations and the requirement of the nilpotency of the BRST
operator \reff{BRST-charge} we have
$$
\be = \frac{25-D}{48}.
$$
This means that we can construct the gauge invariant lagrangian \reff{<Q>}
in the space-time with dimensionality $D\le25$ for any level. This
statement is valid both for the free string and for the string in the
constant electromagnetic field. Thereby, we can describe the interaction of
high spin fields with the constant electromagnetic field in the space-time
with the non-critical dimensionality. A "charge" for such
possibility will be an increase of a state number at each string level%
\footnote{The state number will be the same as in the case of the
dimensional reduction $D\to D-1$.}.

\section{Conclusion.}

In the given work we have considered the electromagnetic interaction of the
second massive state of the open boson string. According to the general
theory this state contains the massive fields with the spins 3~и~1. Using
the BRST quantization method for the open string  we get the gauge invariant
lagrangian describing the interaction of these fields with the constant
electromagnetic field. From the explicit form of transformations and
lagrangian it follows that the presence of the external constant e/m field
lead to the state mixing at the given level. Most likely, the presence of
the external field will lead to a mixing of states at other mass string
levels as well. However, it is not clear if the states of each particular
level will mixed all together or some clusterization of states will occur.
This question requires a separate study.

The author thanks Yu. M. Zinoviev for the useful discussion and the help
in work.


\appendix
\section{}
Here we display the complete non-diagonal lagrangian describing a
propagation of the fields with spins 3 and 1 in the homogeneous
electromagnetic field without the auxiliary fields
\begin{eqnarray}
\label{Lful}
L&=&3
\bar{H}_{1\alpha \mu }\bar{H}_{1\beta \nu }\bar{H}_{1\gamma
\rho }\bar{H}_{1\delta \lambda }{\cal D}_\delta \Phi _{\alpha \beta
\gamma }{\cal D}_\lambda \bar \Phi _{\mu \nu \rho } \non\\&&{} -9
\bar{H}_{1\alpha \mu }\bar{H}_{1\beta \nu }\bar{H}_{1\gamma
\lambda }\bar{H}_{1\delta \rho }{\cal D}_\delta \Phi _{\alpha \beta
\gamma }{\cal D}_\lambda \bar \Phi _{\mu \nu \rho }-3H_{1\beta \mu
}H_{1\alpha \delta }H_{1\gamma \nu }{\cal D}_{\beta \gamma }^2\bar
A_\alpha
\Phi _{\delta \mu \nu } \non\\&&{} -3
\bar{H}_{1\gamma \nu }\bar{H}_{1\alpha \delta }\bar{H}%
_{1\beta \mu }{\cal D}_{\gamma \beta }^2A_\alpha \bar \Phi _{\delta \mu
\nu
}-\frac 14\bar{H}_{1\beta \mu }\bar{H}_{2\gamma \delta
}\bar{H%
}_{2\alpha \nu }{\cal D}_\gamma h_{\alpha \beta }{\cal D}_\nu \bar
h_{\delta
\mu } \non\\&&{} -\frac 12
\bar{H}_{1\beta \nu }\bar{H}_{1\gamma \mu }\bar{H}_{2\alpha
\delta }{\cal D}_\gamma h_{\alpha \beta }{\cal D}_\nu \bar h_{\delta \mu
}+\frac 12\bar{H}_{1\gamma \nu }\bar{H}_{1\beta \mu
}\bar{H}%
_{2\alpha \delta }{\cal D}_\gamma h_{\alpha \beta }{\cal D}_\nu \bar
h_{\delta \mu } \non\\&&{} +\frac 14H_{1\beta \nu }H_{2\delta \mu
}H_{2\alpha
\gamma
}
{\cal D}_{\delta \gamma }^2\bar h_{\alpha \beta }A_{\mu \nu }+\frac 14
\bar{H}_{1\beta \nu }\bar{H}_{2\delta \mu }\bar{H}_{2\alpha
\gamma }{\cal D}_{\delta \gamma }^2h_{\alpha \beta }\bar A_{\mu \nu }
\non\\&&{} 
+\frac 12H_{1\beta \gamma }H_{2\alpha \mu }H_{1\delta \nu }
{\cal D}_{\delta \gamma }^2\bar h_{\alpha \beta }A_{\mu \nu }-\frac
12H_{2\alpha \mu }H_{1\beta \nu }H_{1\gamma \delta }{\cal D}_{\delta
\gamma
}^2\bar h_{\alpha \beta }A_{\mu \nu } \non\\&&{} -\frac 12
\bar{H}_{1\beta \nu }\bar{H}_{2\alpha \mu }\bar{H}_{1\gamma
\delta }{\cal D}_{\delta \gamma }^2h_{\alpha \beta }\bar A_{\mu \nu
}+\frac
12\bar{H}_{2\alpha \mu }\bar{H}_{1\beta \gamma }\bar{H}%
_{1\delta \nu }{\cal D}_{\delta \gamma }^2h_{\alpha \beta }\bar A_{\mu \nu
}
\non\\&&{} -\frac 14H_{2\alpha \gamma }H_{1\beta \delta }
{\cal D}_{\beta \alpha }^2\bar \varphi h_{\gamma \delta }-\frac
14\bar{H%
}_{2\alpha \gamma }\bar{H}_{1\beta \delta }{\cal D}_{\beta \alpha
}^2\varphi \bar h_{\gamma \delta } 
\\&&{}
 +\frac 14
\bar{H}_{1\beta \nu }\bar{H}_{2\delta \mu }\bar{H}_{2\alpha
\gamma }{\cal D}_{\delta \gamma }^2A_{\alpha \beta }\bar A_{\mu \nu
}-\frac
12\bar{H}_{1\beta \nu }\bar{H}_{2\alpha \mu
}\bar{H}_{1\gamma
\delta }{\cal D}_{\delta \gamma }^2A_{\alpha \beta }\bar A_{\mu \nu }
\non\\&&{} 
+\frac 12
\bar{H}_{2\alpha \mu }\bar{H}_{1\beta \gamma }\bar{H}%
_{1\delta \nu }{\cal D}_{\delta \gamma }^2A_{\alpha \beta }\bar A_{\mu \nu
}-\frac 14H_{2\alpha \gamma }H_{1\beta \delta }{\cal D}_{\beta \alpha
}^2\bar \varphi A_{\gamma \delta } \non\\&&{} -\frac 14
\bar{H}_{2\alpha \gamma }\bar{H}_{1\beta \delta }{\cal D}_{\beta
\alpha }^2\varphi \bar A_{\gamma \delta }-\frac 12H_{1\gamma \delta }
\bar{H}_{1\alpha \beta }{\cal D}_\beta A_\alpha {\cal D}_\delta \bar
A_\gamma  \non\\&&{} +
\bar{H}_{1\alpha \gamma }{\cal D}^2A_\alpha \bar A_\gamma +\frac 12
\bar{H}_{1\beta \delta }\bar{H}_{3\alpha \gamma }{\cal D}_\beta
b_\alpha {\cal D}_\delta \bar b_\gamma -\frac 16\bar{H}_{3\alpha
\delta
}\bar{H}_{3\beta \gamma }{\cal D}_\beta b_\alpha {\cal D}_\delta \bar
b_\gamma  \non\\&&{} +\frac 14H_{1\alpha \beta }{\cal D}_{\beta \alpha
}^2\varphi
\bar \varphi -\frac 34\bar{H}_{2\alpha \gamma }\bar{H}_{1\beta
\delta }\bar{H}_{1\mu \nu }\bar{H}_{1\mu \rho }{\cal D}_\gamma
h_{\alpha \beta }\bar \Phi _{\delta \nu \rho }i
\non\\&&{}
+\frac 34H_{2\alpha \gamma }H_{1\beta \delta }
\bar{H}_{1\mu \nu }\bar{H}_{1\mu \rho }{\cal D}_\gamma \bar
h_{\alpha \beta }\Phi _{\delta \nu \rho }i+3H_{1\beta \mu }H_{1\gamma \nu
}H_{2\alpha \delta }H_{1\delta \rho }{\cal D}_\gamma \bar h_{\alpha \beta
}\Phi _{\mu \nu \rho }i \non\\&&{} -3
\bar{H}_{1\delta \rho }\bar{H}_{1\gamma \nu
}\bar{H}_{1\beta
\mu }\bar{H}_{2\alpha \delta }{\cal D}_\gamma h_{\alpha \beta }\bar
\Phi _{\mu \nu \rho }i+\frac 34H_{2\alpha \gamma }H_{1\beta \delta }
\bar{H}_{1\mu \nu }\bar{H}_{1\mu \rho }{\cal D}_\gamma \bar
A_{\alpha \beta }\Phi _{\delta \nu \rho }i \non\\&&{} -\frac 34
\bar{H}_{2\alpha \gamma }\bar{H}_{1\beta \delta }\bar{H}%
_{1\mu \nu }\bar{H}_{1\mu \rho }{\cal D}_\gamma A_{\alpha \beta }\bar
\Phi _{\delta \nu \rho }i+3H_{1\beta \mu }H_{1\gamma \nu }H_{2\alpha
\delta
}H_{1\delta \rho }{\cal D}_\gamma \bar A_{\alpha \beta }\Phi _{\mu \nu
\rho
}i \non\\&&{} -3
\bar{H}_{1\delta \rho }\bar{H}_{1\gamma \nu
}\bar{H}_{1\beta
\mu }\bar{H}_{2\alpha \delta }{\cal D}_\gamma A_{\alpha \beta }\bar
\Phi _{\mu \nu \rho }i-\frac 34H_{1\alpha \beta }\bar{H}_{1\gamma \mu
}
\bar{H}_{1\gamma \delta }{\cal D}_\alpha \bar \varphi \Phi _{\beta
\delta \mu }i \non\\&&{} +\frac 34
\bar{H}_{1\gamma \mu }\bar{H}_{1\gamma \delta }\bar{H}%
_{1\alpha \beta }{\cal D}_\alpha \varphi \bar \Phi _{\beta \delta \mu
}i+\frac 14H_{1\gamma \mu }H_{2\beta \delta }H_{3\alpha \gamma }{\cal D}%
_\beta \bar b_\alpha h_{\delta \mu }i \non\\&&{} -\frac 14
\bar{H}_{1\gamma \mu }\bar{H}_{3\alpha \gamma }\bar{H}%
_{2\beta \delta }{\cal D}_\beta b_\alpha \bar h_{\delta \mu }i+\frac
12H_{1\beta \mu }H_{2\gamma \delta }H_{3\alpha \gamma }{\cal D}_\beta \bar
b_\alpha h_{\delta \mu }i \non\\&&{} -\frac 12
\bar{H}_{1\beta \mu }\bar{H}_{2\gamma \delta }\bar{H}%
_{3\alpha \gamma }{\cal D}_\beta b_\alpha \bar h_{\delta \mu }i+\frac
16H_{1\gamma \mu }H_{2\gamma \delta }H_{3\alpha \beta }{\cal D}_\beta \bar
b_\alpha h_{\delta \mu }i \non\\&&{} -\frac 16
\bar{H}_{1\gamma \mu }\bar{H}_{2\gamma \delta }\bar{H}%
_{3\alpha \beta }{\cal D}_\beta b_\alpha \bar h_{\delta \mu }i-\frac
34H_{1\alpha \delta }H_{2\beta \gamma }{\cal D}_\beta \bar A_\alpha
h_{\gamma \delta }i \non\\&&{} +\frac 34
\bar{H}_{1\alpha \delta }\bar{H}_{2\beta \gamma }{\cal D}_\beta
A_\alpha \bar h_{\gamma \delta }i-\frac 12H_{1\beta \gamma }H_{1\alpha \mu
}H_{2\gamma \delta }{\cal D}_\beta \bar A_\alpha h_{\delta \mu }i
\non\\&&{}
+\frac
12
\bar{H}_{1\alpha \mu }\bar{H}_{2\gamma \delta }\bar{H}%
_{1\beta \gamma }{\cal D}_\beta A_\alpha \bar h_{\delta \mu }i-H_{1\beta
\mu
}H_{2\gamma \delta }H_{1\alpha \gamma }{\cal D}_\beta \bar A_\alpha
h_{\delta \mu }i \non\\&&{} +
\bar{H}_{1\beta \mu }\bar{H}_{2\gamma \delta }\bar{H}%
_{1\alpha \gamma }{\cal D}_\beta A_\alpha \bar h_{\delta \mu }i+\frac
14H_{1\gamma \mu }H_{2\beta \delta }H_{3\alpha \gamma }{\cal D}_\beta \bar
b_\alpha A_{\delta \mu }i \non\\&&{} -\frac 14
\bar{H}_{1\gamma \mu }\bar{H}_{3\alpha \gamma }\bar{H}%
_{2\beta \delta }{\cal D}_\beta b_\alpha \bar A_{\delta \mu }i+\frac
12H_{1\beta \mu }H_{2\gamma \delta }H_{3\alpha \gamma }{\cal D}_\beta \bar
b_\alpha A_{\delta \mu }i \non\\&&{} -\frac 12
\bar{H}_{1\beta \mu }\bar{H}_{2\gamma \delta }\bar{H}%
_{3\alpha \gamma }{\cal D}_\beta b_\alpha \bar A_{\delta \mu }i+\frac
16H_{1\gamma \mu }H_{2\gamma \delta }H_{3\alpha \beta }{\cal D}_\beta \bar
b_\alpha A_{\delta \mu }i \non\\&&{} -\frac 16
\bar{H}_{1\gamma \mu }\bar{H}_{2\gamma \delta }\bar{H}%
_{3\alpha \beta }{\cal D}_\beta b_\alpha \bar A_{\delta \mu }i-\frac
34H_{1\alpha \delta }H_{2\beta \gamma }{\cal D}_\beta \bar A_\alpha
A_{\gamma \delta }i \non\\&&{} +\frac 34\bar{H}_{1\alpha \delta
}\bar{H}%
_{2\beta \gamma }{\cal D}_\beta A_\alpha \bar A_{\gamma \delta }i-\frac
12H_{1\beta \gamma }H_{1\alpha \mu }H_{2\gamma \delta }{\cal D}_\beta \bar
A_\alpha A_{\delta \mu }i
\non\\&&{}
+\frac 12 
\bar{H}_{1\alpha \mu }\bar{H}_{2\gamma \delta }\bar{H}%
_{1\beta \gamma }{\cal D}_\beta A_\alpha \bar A_{\delta \mu }i-H_{1\beta
\mu
}H_{2\gamma \delta }H_{1\alpha \gamma }{\cal D}_\beta \bar A_\alpha
A_{\delta \mu }i \non\\&&{} + 
\bar{H}_{1\beta \mu }\bar{H}_{2\gamma \delta }\bar{H}%
_{1\alpha \gamma }{\cal D}_\beta A_\alpha \bar A_{\delta \mu }i-\frac
23H_{3\alpha \beta }{\cal D}_\alpha \bar \varphi b_\beta i+\frac
23\bar{%
H}_{3\alpha \beta }{\cal D}_\alpha \varphi \bar b_\beta i \non\\&&{} -\frac
14H_{1\alpha \beta }H_{3\beta \gamma } 
{\cal D}_\alpha \bar \varphi b_\gamma i+\frac 14\bar{H}_{3\beta
\gamma
} \bar{H}_{1\alpha \beta }{\cal D}_\alpha \varphi \bar b_\gamma
i+\frac
34H_{1\alpha \beta }{\cal D}_\alpha \bar \varphi A_\beta i \non\\&&{}
-\frac
34 
\bar{H}_{1\alpha \beta }{\cal D}_\alpha \varphi \bar A_\beta i-6 
\bar{H}_{1\gamma \delta }\bar{H}_{1\alpha \beta }\bar{H}%
_{1\mu \nu }\Phi _{\alpha \gamma \mu }\bar \Phi _{\beta \delta \nu }
\non\\&&{} 
+18H_{1\alpha \beta } 
\bar{H}_{1\alpha \gamma }\bar{H}_{1\delta \mu
}\bar{H}_{1\nu
\rho }\Phi _{\beta \delta \nu }\bar \Phi _{\gamma \mu \rho } \non\\&&{}
-\frac
94 
\bar{H}_{1\delta \nu }\bar{H}_{1\alpha \gamma }\bar{H}%
_{1\alpha \beta }\bar{H}_{1\delta \mu }\bar{H}_{1\rho \lambda
}\Phi _{\beta \gamma \rho }\bar \Phi _{\mu \nu \lambda } \non\\&&{}
-9F_{\beta
\lambda }H_{1\alpha \beta } 
\bar{H}_{1\gamma \delta }\bar{H}_{1\mu \nu }\bar{H}_{1\rho
\lambda }\Phi _{\gamma \mu \rho }\bar \Phi _{\alpha \delta \nu } \non\\&&{}
-\frac
34H_{1\alpha \beta } 
\bar{H}_{1\delta \nu }\bar{H}_{3\alpha \gamma }\bar{H}%
_{1\delta \mu }\Phi _{\beta \mu \nu }\bar b_\gamma -\frac 34H_{3\alpha
\beta
}\bar{H}_{1\delta \nu }\bar{H}_{1\alpha \gamma }\bar{H}%
_{1\delta \mu }\bar \Phi _{\gamma \mu \nu }b_\beta \non\\&&{} +\frac
94H_{1\alpha
\beta } 
\bar{H}_{1\gamma \mu }\bar{H}_{1\gamma \delta }\bar \Phi
_{\alpha
\delta \mu }A_\beta +\frac 94\bar{H}_{1\alpha \gamma }\bar{H}%
_{1\alpha \beta }\bar{H}_{1\delta \mu }\Phi _{\beta \gamma \delta
}\bar
A_\mu \non\\&&{} +H_{2\alpha \beta } 
\bar{H}_{2\alpha \gamma }\bar{H}_{1\delta \mu }h_{\beta \delta
}\bar h_{\gamma \mu }+F_{\gamma \mu }\bar{H}_{1\gamma \delta
}\bar{%
H}_{2\alpha \beta }h_{\alpha \mu }\bar h_{\beta \delta } \non\\&&{} -\frac
12H_{1\gamma \delta }H_{2\alpha \beta } 
\bar{H}_{1\alpha \nu }\bar{H}_{2\gamma \mu }h_{\beta \delta
}\bar
h_{\mu \nu }-\frac 12H_{2\alpha \beta }\bar{H}_{2\gamma \delta } 
\bar{H}_{1\alpha \gamma }\bar{H}_{1\mu \nu }h_{\beta \mu }\bar
h_{\delta \nu } \non\\&&{} -\frac 16H_{2\alpha \beta }H_{1\alpha \gamma } 
\bar{H}_{2\delta \mu }\bar{H}_{1\delta \nu }h_{\beta \gamma
}\bar
h_{\mu \nu }+\frac 14F_{\alpha \gamma }H_{2\alpha \beta }\bar{H}%
_{2\gamma \delta }\bar{H}_{1\mu \nu }h_{\beta \mu }\bar h_{\delta \nu
}
\non\\&&{} +\frac 12F_{\alpha \mu }H_{1\alpha \beta } 
\bar{H}_{2\gamma \delta }\bar{H}_{1\mu \nu }h_{\beta \gamma
}\bar
h_{\delta \nu }+F_{\gamma \mu }\bar{H}_{1\gamma \delta }\bar{H}%
_{2\alpha \beta }\bar h_{\beta \delta }A_{\alpha \mu } \non\\&&{}
+F_{\delta
\mu
}H_{1\gamma \delta }H_{2\alpha \beta }h_{\beta \mu }\bar A_{\alpha \gamma
}-H_{1\gamma \delta }H_{2\alpha \beta } 
\bar{H}_{2\alpha \mu }h_{\beta \delta }\bar A_{\gamma \mu } \non\\&&{} 
+H_{2\alpha \beta }\bar{H}_{2\alpha \gamma }\bar{H}_{1\delta \mu
}\bar h_{\gamma \mu }A_{\beta \delta }-\frac 12H_{1\gamma \delta
}H_{2\alpha
\beta }\bar{H}_{1\alpha \nu }\bar{H}_{2\gamma \mu }h_{\beta
\delta
}\bar A_{\mu \nu } 
\non\\&&{}
-\frac 12H_{1\gamma \delta }H_{2\alpha \beta } 
\bar{H}_{1\alpha \nu }\bar{H}_{2\gamma \mu }\bar h_{\mu \nu
}A_{\beta \delta }+\frac 12H_{2\alpha \beta }H_{1\delta \mu }\bar{H}%
_{1\alpha \gamma }\bar{H}_{2\gamma \nu }h_{\beta \mu }\bar A_{\delta
\nu } \non\\&&{} -\frac 12H_{2\alpha \beta } 
\bar{H}_{2\gamma \delta }\bar{H}_{1\alpha \gamma }\bar{H}%
_{1\mu \nu }\bar h_{\delta \nu }A_{\beta \mu }-\frac 16H_{2\alpha \beta
}H_{1\alpha \gamma }\bar{H}_{2\delta \mu }\bar{H}_{1\delta \nu
}h_{\beta \gamma }\bar A_{\mu \nu } \non\\&&{} -\frac 16H_{2\alpha \beta
}H_{1\alpha
\gamma } 
\bar{H}_{2\delta \mu }\bar{H}_{1\delta \nu }\bar h_{\mu \nu
}A_{\beta \gamma }+\frac 23H_{2\alpha \beta }H_{1\alpha \gamma }h_{\beta
\gamma }\bar \varphi \non\\&&{} +\frac 23 
\bar{H}_{1\alpha \gamma }\bar{H}_{2\alpha \beta }\bar h_{\beta
\gamma }\varphi -\frac 14F_{\alpha \gamma }H_{1\gamma \delta }H_{2\alpha
\beta }h_{\beta \delta }\bar \varphi +\frac 14F_{\alpha \gamma
}\bar{H}%
_{1\gamma \delta }\bar{H}_{2\alpha \beta }\bar h_{\beta \delta
}\varphi
\non\\&&{} +H_{2\alpha \beta } 
\bar{H}_{2\alpha \gamma }\bar{H}_{1\delta \mu }A_{\beta \delta
}\bar A_{\gamma \mu }-F_{\beta \mu }H_{1\alpha \beta
}\bar{H}_{2\gamma
\delta }A_{\gamma \mu }\bar A_{\alpha \delta } \non\\&&{} -\frac
12H_{1\gamma
\delta
}H_{2\alpha \beta } 
\bar{H}_{1\alpha \nu }\bar{H}_{2\gamma \mu }A_{\beta \delta
}\bar
A_{\mu \nu }-\frac 12H_{2\alpha \beta }\bar{H}_{2\gamma \delta } 
\bar{H}_{1\alpha \gamma }\bar{H}_{1\mu \nu }A_{\beta \mu }\bar
A_{\delta \nu } \non\\&&{} -\frac 16H_{2\alpha \beta }H_{1\alpha \gamma } 
\bar{H}_{2\delta \mu }\bar{H}_{1\delta \nu }A_{\beta \gamma
}\bar
A_{\mu \nu }+\frac 23H_{2\alpha \beta }H_{1\alpha \gamma }A_{\beta \gamma
}\bar \varphi \non\\&&{} +\frac 23 
\bar{H}_{1\alpha \gamma }\bar{H}_{2\alpha \beta }\bar A_{\beta
\gamma }\varphi +\frac 14F_{\alpha \gamma }\bar{H}_{1\gamma \delta } 
\bar{H}_{2\alpha \beta }\bar A_{\beta \delta }\varphi \non\\&&{} -\frac
14F_{\beta \delta } 
\bar{H}_{1\gamma \delta }\bar{H}_{2\alpha \beta }A_{\alpha
\gamma
}\bar \varphi +H_{2\alpha \beta }\bar{H}_{3\alpha \gamma }b_\beta
\bar
b_\gamma -\frac 14H_{3\alpha \beta }\bar{H}_{1\alpha \gamma
}\bar{H%
}_{3\gamma \delta }b_\beta \bar b_\delta \non\\&&{} -\frac 12H_{3\alpha
\beta
} 
\bar{H}_{2\alpha \gamma }\bar{H}_{3\gamma \delta }b_\beta \bar
b_\delta +\frac 16F_{\alpha \gamma }H_{3\alpha \beta
}\bar{H}_{3\gamma
\delta }b_\beta \bar b_\delta \non\\&&{} +\frac 34H_{1\alpha \beta } 
\bar{H}_{3\alpha \gamma }\bar b_\gamma A_\beta +\frac 34H_{3\alpha
\beta }\bar{H}_{1\alpha \gamma }b_\beta \bar A_\gamma +\frac
12H_{1\beta \gamma }H_{2\alpha \beta }\bar{H}_{3\alpha \delta }\bar
b_\delta A_\gamma \non\\&&{} +\frac 12H_{3\alpha \beta } 
\bar{H}_{2\alpha \gamma }\bar{H}_{1\gamma \delta }b_\beta \bar
A_\delta -\frac 94\bar{H}_{1\alpha \beta }A_\alpha \bar A_\beta
-H_{2\alpha \beta }\bar{H}_{1\alpha \gamma }A_\beta \bar A_\gamma
\non\\&&{} 
-\frac 12F_{\alpha \beta }^2 
\bar{H}_{1\gamma \delta }A_\gamma \bar A_\delta -\frac 12F_{\alpha
\gamma }H_{1\alpha \beta }\bar{H}_{1\gamma \delta }A_\beta \bar
A_\delta 
\non\\\non&&{}
 -\frac 12H_{1\beta \gamma }H_{2\alpha \beta }\bar{H}%
_{1\alpha \delta }A_\gamma \bar A_\delta -\frac 83\varphi \bar \varphi
-\frac 14F_{\alpha \beta }^2\varphi \bar \varphi 
\end{eqnarray}

\end{document}